\documentstyle[twoside,fleqn,espcrc2,epsf,psfig,epsfig]{article}

\title{Lyapunov exponents in Minkowskian U(1) gauge theory
\thanks{Support by FWF P14435-TPH and computer code for Lyapunov exponents
by T.S. Bir\'o are gratefully acknowledged.}}

\author{Harald Markum, Rainer Pullirsch, Wolfgang Sakuler
\address{Atominstitut, Technische Universit\"at Wien, A-1040 Vienna, Austria}}

\begin{document}

\begin{abstract} 
        U(1) gauge fields are decomposed into a monopole and photon part
        across the phase transition from the confinement to the Coulomb
        phase.  We analyze the leading Lyapunov exponents of such
  	gauge field configurations on the lattice
        which are initialized by quantum Monte Carlo simulations.
        We observe that the monopole field carries the same
        Lyapunov exponent as the original U(1) field.
        Evidence is found that monopole fields stay chaotic in the 
        continuum whereas the photon fields are regular.
\end{abstract}
\maketitle

\section {Monopole and photon part of U(1)}

We begin with a $4d$ U(1) gauge theory described by the 
Euclidean action
$ S \lbrace U_l \rbrace = \beta \: \sum_p (1 - \cos \theta_p )$,
where $U_l = U_{x,\mu} = \exp (i\theta_{x,\mu}) $ and
$
  \theta_p =
 \theta_{x,\mu} +
 \theta_{x+\hat{\mu},\nu} -
 \theta_{x+\hat{\nu},\mu} -
 \theta_{x,\nu}\ . $
We are interested in the relationship between monopoles 
and classical chaos across the phase transition
at $\beta_c \approx 1.01$. 
Following Ref.~\cite{StWe92}, we have
factorized our gauge configurations
into monopole and photon fields.
The U(1) plaquette angles $\theta_{x,\mu\nu}$ are decomposed into the
``physical'' electromagnetic flux through the plaquette
$\bar \theta_{x,\mu\nu}$ and a number $m_{x,\mu\nu}$ of Dirac strings
through the plaquette
\begin{equation} \label{Dirac_string_def}
 \theta_{x,\mu\nu} = \bar \theta_{x,\mu\nu} + 2\pi\,m_{x,\mu\nu}\ ,
\end{equation}
where $\bar \theta_{x,\mu\nu}\in (-\pi,+\pi]$ and
$m_{x,\mu\nu} \ne 0$ is called a Dirac plaquette.

\section{Classical chaotic dynamics from quantum Monte Carlo
         initial states}

Chaotic dynamics in general is characterized by the
spectrum of Lyapunov exponents. These exponents, if they are positive,
reflect an exponential divergence of initially adjacent configurations.
In case of symmetries inherent in the Hamiltonian of the system
there are corresponding zero values of these exponents. Finally
negative exponents belong to irrelevant directions in the phase
space: perturbation components in these directions die out
exponentially. Pure gauge fields on the lattice show a characteristic
Lyapunov spectrum consisting of one third of each kind of
exponents~\cite{BOOK}.
Assuming this general structure of the Lyapunov spectrum we
investigate presently its magnitude only, namely the maximal
value of the Lyapunov exponent, $L_{{\rm max}}$.

\begin{figure*}[ht]
\centerline{{\hspace*{5mm}\psfig{figure=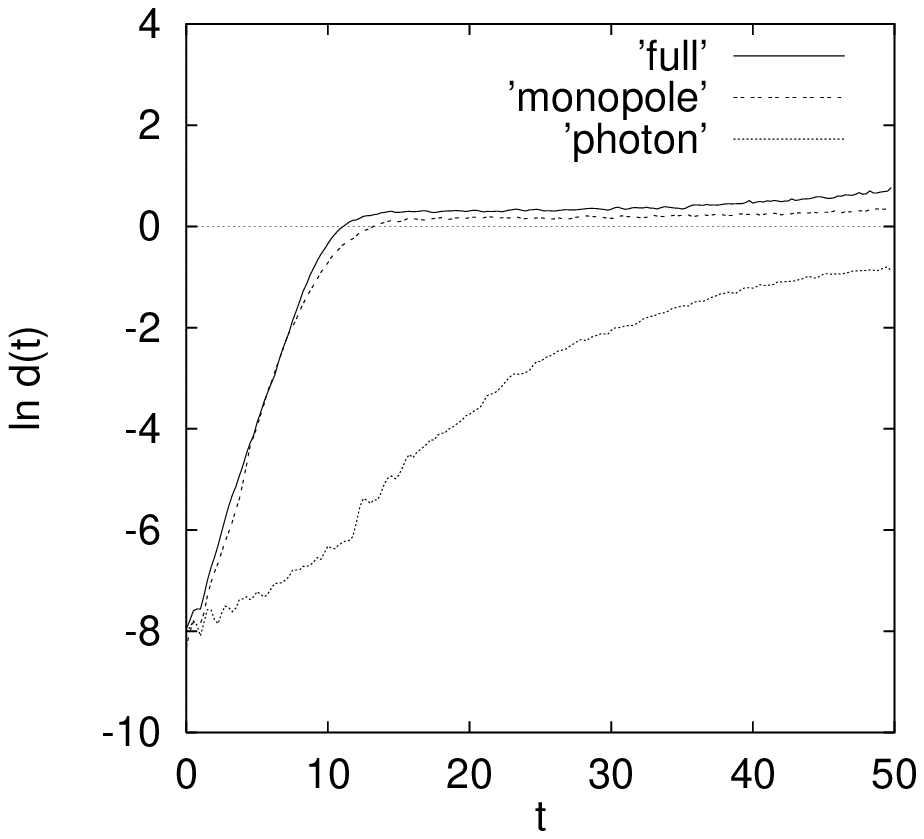,width=6cm}}\hspace{5mm}
{\psfig{figure=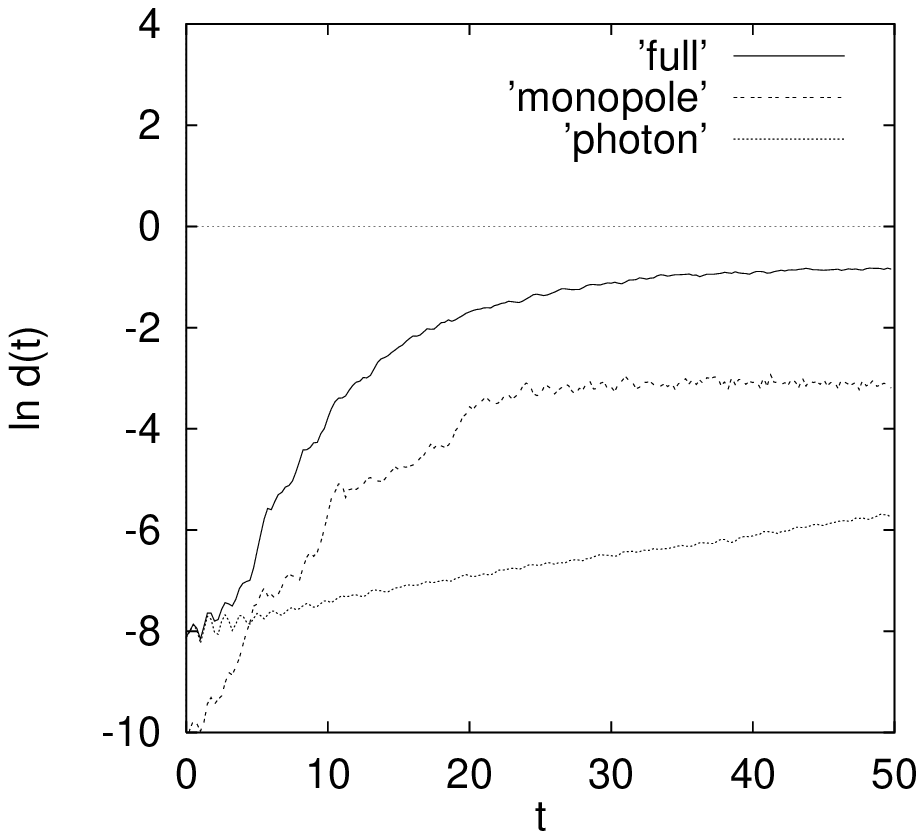,width=6cm}}}
\vspace*{-8mm}
\caption{
  Exponentially diverging distance in real time of initially adjacent U(1) field
  configurations on a $12^3$ lattice prepared at $\beta=0.9$ in the
  confinement (left) and at $\beta=1.1$ in the Coulomb
  phase (right).
\vspace*{-5mm}
\label{Fig2}
 }
\end{figure*}
The general definition of the Lyapunov exponent is based on a
distance measure $d(t)$ in phase space,
\begin{equation}
L := \lim_{t\rightarrow\infty} \lim_{d(0)\rightarrow 0}
\frac{1}{t} \ln \frac{d(t)}{d(0)}.
\end{equation}
In case of conservative dynamics the sum of all Lyapunov exponents
is zero according to Liouville's theorem, $\sum L_i = 0$.
We utilize the gauge invariant distance measure consisting of
the local differences of energy densities between two $3d$ field configurations
on the lattice:
\begin{equation}
d : = \frac{1}{N_P} \sum_P\nolimits \, \left| {\rm tr} U_P - {\rm tr} U'_P \right|.
\end{equation}
Here the symbol $\sum_P$ stands for the sum over all $N_P$ plaquettes,
so this distance is bound in the interval $(0,2N)$ for the group
SU(N). $U_P$ and $U'_P$ are the familiar plaquette variables, constructed from
the basic link variables $U_{x,i}$,
\begin{equation}
U_{x,i} = \exp \left( aA_{x,i}^cT^c \right)\: ,
\end{equation}
located on lattice links pointing from the position $x=(x_1,x_2,x_3)$ to
$x+ae_i$. The generators of the group are
$T^c = -ig\tau^c/2$ with $\tau^c$ being the Pauli matrices
in case of SU(2) and $A_{x,i}^c$ is the vector potential.
The elementary plaquette variable is constructed for a plaquette with a
corner at $x$ and lying in the $ij$-plane as
$U_{x,ij} = U_{x,i} U_{x+i,j} U^{\dag}_{x+j,i} U^{\dag}_{x,j}$.
It is related to the magnetic field strength $B_{x,k}^c$:
\begin{equation}
U_{x,ij} = \exp \left( \varepsilon_{ijk} a B_{x,k}^c T^c \right).
\end{equation}
The electric field strength $E_{x,i}^c$ is related to the canonically conjugate
momentum $P_{x,i} = \dot{U}_{x,i}$ via
\begin{equation}
E^c_{x,i} = \frac{2a}{g^3} {\rm tr} \left( T^c \dot{U}_{x,i} U^{\dag}_{x,i} \right).
\end{equation}

The Hamiltonian of the lattice gauge field system can be casted into
the form
\begin{equation}
H = \sum \left[ \frac{1}{2} \langle P, P \rangle \, + \,
 1 - \frac{1}{4} \langle U, V \rangle \right].
\end{equation}
Here the scalar product stands for
$\langle A, B \rangle = \frac{1}{2} {\rm tr} (A B^{\dag} )$.
The staple variable $V$ is a sum of triple products of elementary
link variables closing a plaquette with the chosen link $U$.
This way the Hamiltonian is formally written as a sum over link
contributions and $V$ plays the role of the classical force
acting on the link variable $U$. 

We prepare the initial field configurations
from a standard four dimensional Euclidean Monte Carlo program on
a $12^3\times 4$ lattice varying the inverse gauge coupling $\beta$~\cite{SU2}.
We relate such four dimensional Euclidean
lattice field configurations to Minkow\-skian momenta and fields
for the three dimensional Hamiltonian simulation
by identifying a fixed time slice of the four dimensional lattice.

\section{Chaos, confinement and continuum }

We start the presentation of our results with a characteristic example
of the time evolution of the distance between initially adjacent
configurations. An initial state prepared by a standard four dimensional
Monte Carlo simulation is evolved according to the classical Hamiltonian dynamics
in real time. Afterwards this initial state is rotated locally by
group elements which are chosen randomly near to the unity.
The time evolution of this slightly rotated configuration is then
pursued and finally the distance between these two evolutions
is calculated at the corresponding times.
A typical exponential rise of this distance followed by a saturation
can be inspected in Fig.~\ref{Fig2} from an example of U(1) gauge theory
in the confinement phase and in the Coulomb phase.
While the saturation is an artifact of
the compact distance measure of the lattice, the exponential rise
(the linear rise of the logarithm)
can be used for the determination of the leading Lyapunov exponent.
The left plot exhibits that in the confinement phase the original
field and its monopole part have similar Lyapunov exponents whereas
the photon part has a smaller $L_{max}$. The right plot in the Coulomb
phase suggests that all slopes and consequently the Lyapunov
exponents of all fields decrease substantially.

\begin{figure}[t]
\centerline{\psfig{figure=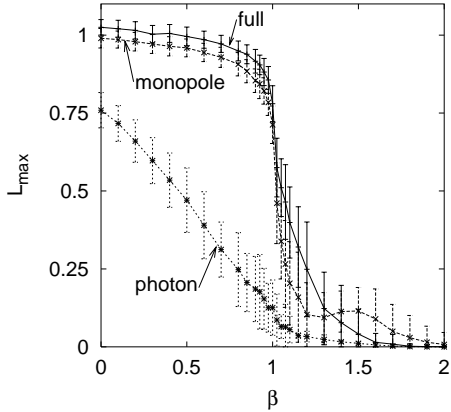,width=6cm}}
\vspace*{-8mm}
\caption{Lyapunov exponent of the decomposed U(1) fields as a function
         of coupling.
\vspace*{-5mm}
\label{Fig3}
 }
\end{figure}
\begin{figure}[t]
\centerline{\psfig{figure=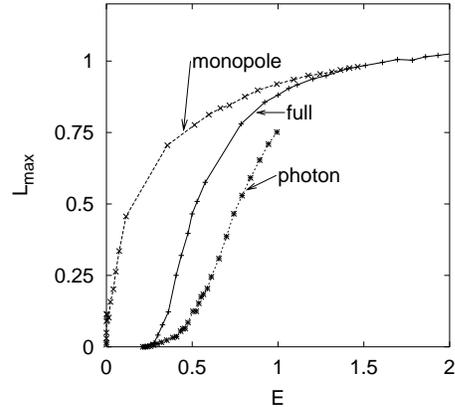,width=6cm}}
\vspace*{-8mm}
\caption{
  Comparison of average maximal Lyapunov exponents as a function of the
  scaled average energy per plaquette $ag^2E$. The full U(1) theory
  shows an approximately quadratic behavior in the weak coupling regime
  which is more pronounced for the photon field. The monopole field 
  indicates a linear relation.
\vspace*{-5mm}
\label{Fig4}
 }
\end{figure}
The main result of the present study is the dependence of the leading
Lyapunov exponent $L_{{\rm max}}$ on the inverse coupling strength $\beta$,
displayed in Fig.~\ref{Fig3}.
As expected the strong coupling phase
is more chaotic.  The transition reflects
the critical coupling to the Coulomb phase. 
The plot shows that the monopole fields carry Lyapunov
exponents of nearly the same size as the full U(1) fields. The photon
fields yield a non-vanishing value in the confinement ascending toward $\beta=0$
for randomized fields which indicates that the decomposition
works well only around the phase transition.

An interesting result concerning the continuum limit can be viewed from Fig.~\ref{Fig4}
which shows the energy dependence of the Lyapunov exponents for the U(1) theory and its
components. One observes an approximately linear relation for the monopole part while a
quadratic relation is suggested for the photon part in the weak coupling regime.
From scaling arguments one expects a functional relationship between
the Lyapunov exponent and the energy \cite{BOOK,SCALING}
\begin{equation}
L(a) \propto a^{k-1} E^{k}(a) ,
\label{scaling}
\end{equation}
with the exponent $k$ being crucial for the continuum limit of the
classical field theory. A value of $k < 1$ leads to a
divergent Lyapunov exponent, while $k > 1$ yields a vanishing $L$ in
the continuum. The case $k = 1$ is special leading to a finite non-zero
Lyapunov exponent. Our analysis of the scaling relation (\ref{scaling})
gives evidence, that the classical compact U(1) lattice gauge theory
and especially the photon field have $k \approx 2$ and with $L(a) \to 0$
a regular continuum theory. The monopole field signals $k \approx 1$ and
stays chaotic approaching the continuum.

\end{document}